\documentclass[amscls]{revtex4}
\newcommand{\be}{\begin{equation}}
\newcommand{\ee}{\end{equation}}
\newcommand{\ba}{\begin{eqnarray}}
\newcommand{\ea}{\end{eqnarray}}
\begin{document}

\title{Near-Horizon Geometry and the Entropy of a Minimally
Coupled Scalar Field in the Kerr Black Hole}

\author{Kaushik Ghosh\footnote{E-mail ghosh{\_}kaushik06@yahoo.co.in}}
\affiliation{Vivekananda College,University of Calcutta, 
269, Diamond Harbour Road, Kolkata - 700063, India}

\maketitle

\section*{Abstract}

In this article we will discuss a Lorentzian sector calculation of the entropy of a minimally
coupled scalar field in the Kerr black hole background. We will use the brick wall model of
’t Hooft. In the Kerr black hole, complications arise due to the absence of a global timelike
Killing field and the presence of the ergosphere. Nevertheless, it is possible to calculate the
entropy of a thin shell of matter field in the near-horizon region using the brick wall model. The
corresponding leading order entropy of the nonsuperradiant modes is found to be proportional to
the area of the horizon and is logarithmically divergent. Thus, to the leading order, the entropy of a three-dimensional
system in the near-horizon region is proportional to the boundary surface.
This aspect is also valid in the Schwarzschild black hole. The corresponding internal energy remains finite if the
entropy is chosen to be of the order of the black hole entropy itself.
For a fixed value of the brick wall cut-off, the leading order entropy in the Kerr black hole is found
to be half of the corresponding term in the Schwarzschild black hole. This is due to rotation and is
consistent with the preferential emission of particles in the Kerr black hole with azimuthal angular momentum in
the same direction as that of the black hole itself. However, we can obtain the Schwarzschild case expression
by including a subleading term and taking the appropriate slow rotation limit. In this version, we will discuss a sub-leading term for the scalar field entropy in Schwarzschild black hole. This term reduces to an expression similar to the corresponding flat spacetime entropy in the limit when radius of event horizon is vanishing.

Key-words: Kerr black hole, scalar field entropy, brick wall model, area law, logarithmic divergence

PACS numbers: 97.60.Lf, 04.70.Dy, 04.62.+v, 11.10.Wx 	



\newpage

\section {I. Introduction}
 
Since the four laws of black hole mechanics were formulated [1], there has been much effort to relate
the laws of black hole mechanics to those of thermodynamics. The area theorem led Bekenstein to
assign an entropy to a black hole [2]. He proposed a generalized second law of thermodynamics by
considering the sum of a multiple of the area of horizon and the entropy of ordinary matter moving
in the black hole background. A black hole can be assigned a temperature equal to a multiple of the
surface gravity of the horizon. However, unless the black hole can emit particles, the second law can
be violated by considering a black body radiation surrounding the black hole at a lower temperature.
Hawking established the thermodynamical aspects of black holes by showing that a black hole can
radiate like a hot body at a temperature equal to a multiple of the surface gravity of the horizon [3].
The entropy of a black hole, considered as a thermodynamical system, was found to be
${A\over{4}}$. Here $A$ is the area of the horizon.

We have to consider quantum field theory in curved spaces in most of the works related to black
hole thermodynamics. This is a nontrivial issue since, in a curved space, the vacuum and finite particle
states are dependent on the observer. We do not face this problem for quantum field theory in flat space
due to Poincaré invariance. However, even in flat space, a uniformly accelerated observer detects a
thermal spectrum when the field is in the Minkowski vacuum [4,5]. The temperature is dependent
on the proper acceleration of the observer. A corresponding situation arises in the Schwarzschild
black hole for a static observer outside the event horizon. The temperature is the same as that of the
black hole if the static observer is at a large distance from the horizon. The corresponding vacuum
is the Hartle–Hawking vacuum [6]. G. ’t Hooft proposed a model to calculate the entropy of a scalar
field in the Schwarzschild black hole in thermal equilibrium with the black hole [7]. He used the
Wentzel–Kramers–Brillouin (WKB) approximation to count the states. We can define a thermal
equilibrium between a black hole and black body radiation surrounding the black hole at the same
temperature as that of the black hole. The amount of radiation absorbed and emitted by the black
hole is the same. This is an unstable equilibrium due to the negative specific heat of the black hole,
which is characteristic of self-gravitating systems [8]. This is discussed for a few systems in Ref. [3].
A small perturbation in the background can give rise to a runaway situation. Divergences appear
when we try to calculate the entropy of the scalar field [7]. This is associated with the continuous
energy spectrum and large values of the angular quantum numbers of the scalar wave equation. To
regulate the divergence, ’t Hooft proposed a boundary condition on the scalar field near the horizon.
He assumed the scalar field to be vanishing at a small distance away from the horizon. This radial
parameter is known as the brick wall cut-off parameter. This boundary condition is a good model
since in thermodynamic equilibrium there is no net interchange of particles between the black hole
and the surrounding matter. The boundary condition is similar to a perfectly reflecting mirror. An
observer associated with the horizon detects no particle [6] and the brick wall boundary condition is
reasonable in this respect. We can easily construct solutions of the scalar wave equation that vanish
at the brick wall [9]. These are similar to the solutions having zero Cauchy data on the horizon and
are considered in Hawking radiation. This will be discussed for the massless field in Sect. 2.
The brick wall boundary condition is also discussed in detail by ’t Hooft [7]. The WKB quantization
rule together with the brick wall boundary condition introduce the brick wall cut-off parameter as a
regularizing parameter on the otherwise continuous energy spectrum of the scalar field wave equation.
This is an advantage of the brick wall model where a mathematical cut-off can be directly expressed
in terms of a physically relevant parameter. The expression of entropy obtained earlier was linearly
divergent in the brick wall cut-off parameter. However, the metric component $g_{rr}$ 
has a simple pole at
the horizon. The WKB quantization rule indicates that the divergence is expected to be logarithmic. In
the Schwarzschild black hole, we can also use the blueshift factor together with the form of the proper
distance in the near-horizon region to understand this logarithmic divergence. This is also evident
from the nature of the solutions in the near-horizon region and will be discussed in the later sections. A
logarithmically divergent expression for the scalar field entropy in the Bañados–Teitelboim–Zanelli
black hole is obtained in [10] using a different method. Similar behavior is obtained in the optical
metric approach [11].

In a series of recently published articles [12$\textendash$14], we have shown that an improved counting of
the scalar field states leads to a logarithmically divergent expression for the scalar field entropy. We
have found that the WKB approximation is particularly suitable for the scalar field solutions that
are stationary in the near-horizon region, and again at a large distance away from the horizon. This
approximately corresponds to a thin shell of scalar field confined in the near-horizon region due to
the large width of the intermediate potential barrier [15]. We will find that the thickness of the thin
shell imposes an upper limit on the allowed angular quantum numbers. The divergences
can be removed by considering the stress tensor, which is well behaved when the field is in the
Hartle–Hawking vacuum [16]. ’t Hooft removed the divergence by equating the scalar field entropy
to the black hole entropy. When we do so, the internal energy is found to be finite and the brick wall
is almost coincident with the horizon [12]. This is unlike the case in [7], where the brick wall cut-off
is of the order of the Planck length. The leading-order term of the scalar field entropy is
proportional to the area of the horizon. Another interesting aspect of our work is that the proper
distance thickness of the thin shell together with the brick wall cut-off parameter behave like a scale
parameter similar to the Plack length in the gravitational entropy. Thus, the degrees of freedom of a three dimensional thin shell is principally related to the boundary surface area. This is similar to the gravitational entropy of the black hole itself. In section:3, we will discuss a sub-leading term whose form reduces to the familiar expression in flat spacetime in the appropriate flat spacetime limit. Thus, the method used by us to get the thin shell entropy in the near horizon region is consistent with both gravitational blueshift and flat spacetime limit.

A few authors have followed the brick wall model of 't Hooft to use the scalar field entropy for various
purposes; the relevant references will be given in Sect. 3. A more complete set of references can
be found in [12,17]. The present article is also significant in this respect. The scalar field entropy
can be calculated from the Euclidean sector of the black hole [6,17$\textendash$22]. One again considers the
near-horizon region with boundary conditions similar to those considered by us. The scalar field
entropy is again proportional to the horizon surface area. However, the scalar field entropy contains
an undetermined parameter. This parameter can be fixed by comparing with the Lorentzian sector
expression. Thus, the logarithmically divergent expression of Lorentzian sector entropy makes
the Euclidean sector expressions only logarithmically divergent compared to the quadratically and
quartically divergent expressions obtained earlier. These discussions are given in detail in [12]

In the present article we will consider the entropy of a minimally coupled scalar field in the Kerr
black hole. We will consider again a thin shell of scalar field confined in the near horizon region.
We are considering scalar field modes that are stationary in the near-horizon region and again at
a far distance from the horizon. The amplitude at asymptotic infinity is negligible. The Kerr black
hole is stationary and only axisymmetric. There does not exist a global timelike Killing vector field.
The metric component
$g_{rr}$ 
has a simple pole at the outer horizon. Thus the brick wall model is
expected to give a logarithmically divergent expression. This appears to be true, and the scalar field
entropy is again proportional to the surface area of the horizon. The time translation generator
$({{\partial}\over{\partial t}})^{a}$ 
is a Killing field, but unlike the Schwarzschild case it is not globally timelike. This Killing field
becomes spacelike within the ergosphere, which is lying in the near-horizon region surrounding the
black hole. The other Killing field
$({{\partial}\over{\partial \phi}})^{a}$ 
is spacelike everywhere. Here, we have used the notation
of Wald [8]. Since we are considering an observer at a large distance from the horizon, we can use
the Killing field
$({{\partial}\over{\partial t}})^{a}$  
to define energy with respect to that observer. The solutions are taken to be
stationary with respect to 
$t, \phi$. 
In the Kerr black hole we have two sectors of solutions. The first sector
contains the superradiant modes [23,24] and the second sector consists of the ordinary solutions. The
argument in the exponential of the partition function becomes negative for a superradiant mode [3].
The superradiant modes give stimulated radiation. Their contribution to the thermal radiation and
hence to the corresponding entropy is not very significant. The contribution of the ordinary modes
is similar to the Schwarzschild black hole. The corresponding leading-order entropy is proportional
to the area of the event horizon and is logarithmically divergent in the brick wall cut-off parameter.
Although the Killing vector 
$({{\partial}\over{\partial t}})^{a}$ 
is not timelike inside the ergosphere, the ergosphere vanishes at the polar axis. The Killing vector 
$({{\partial}\over{\partial t}})^{a}$ 
remains timelike along the polar axis and is null
on the horizon at 
$\theta = 0, \pi$. 
Thus we can consider a timelike geodesic starting at a suitable position at a large distance away from the horizon and intersecting the horizon at 
$\theta = 0$ or $\pi$ 
without ever entering the ergosphere. We can now apply the blueshift argument to understand the logarithmic divergence. 
There is an important but expected difference
between the scalar field entropy in the Kerr black hole and the Schwarzschild black hole. For a fixed value of the brick wall cut-off, the leading order entropy of a thin shell of scalar field confined in the near horizon region in the Kerr black hole is half of the corresponding expression for the Schwarzschild case. 
This is expected due the preferential
emission of particles in the Kerr black hole with azimuthal angular momentum in the same direction
as that of the black hole itself. The thin shell in thermal equilibrium preferentially contains particles
with azimuthal angular momentum in the same direction as that of the black hole. This is consistent with the
frame dragging effect in the near-horizon region. However, we can obtain the Schwarzschild case expression
by including a subleading term and taking the appropriate slow rotation limit. This is discussed in
detail in Section:II. Another important point to note is that the proper radial thickness of the thin shell is
dependent on the polar angle although the radial coordinate thickness and the entropy is independent of the polar angle.
We will discuss these issues in Sect.II.

The Lorentzian sector calculation of the scalar field entropy in the Kerr black hole is more significant since
the corresponding Euclidean sector literature is not well formulated [20,21]. The expressions
obtained in this article are significant in the context of entanglement entropy and the holographic principles. We
will discuss these aspects and the significance of the results obtained in this article in Sect.III.

\section {Entropy of a Thin Shell of Scalar Field in the Kerr Black Hole}

In this section we will calculate the entropy of a minimally coupled scalar field in the Kerr black hole. We will discuss the case of a massless field but the results can be easily extended for a massive case [14]. We will follow the method discussed in [12,13] and consider a thin shell of scalar field surrounding the horizon.

The space time metric of the Kerr black hole is given by the following expression:

\be 
d{s^2} = -{\Delta \over{\rho^2}}
{[dt - a({{\sin^2}\theta}){d\phi}]^2} +
{{{\sin^2}\theta} \over {\rho}^2}
[(r^2 + a^2){d\phi} - a~dt]^2
+ {{\rho^2} \over {\Delta}}{dr^2}
+ {\rho}^2{d\theta^2} ~.
\ee 

\noindent{ Where,}

\be 
{\rho}^2 = r^2 + {a^2}{{\cos^2}\theta}, ~~~~~
\Delta = r^2 - 2Mr + a^2 ~.
\ee

\noindent{ Here $M$ is the mass and 
$a$ is the angular momentum per unit mass of the black hole.} 

The mass and angular momentum are defined with respect to an asymptotic observer. The position of the horizons are given by the zeros of $\Delta$. There are two horizons which are given by the following expressions:

\be 
r_{\pm} = M {\pm} \sqrt{M^2 - a^2} ~.
\ee

\noindent{ The event horizon is at $r = r_{+}$.}

It is obvious from the line element that the metric is not static but is only stationary and axisymmetric. There are two Killing vectors: $({{\partial}\over{\partial t}})^{a}$ and $({{\partial}\over{\partial \phi}})^{a}$. In terms of the metric in Eqn.(1), they can be expressed as:
${{\xi}_t} = (1,0,0,0)$ and    
${{\xi}_{\phi}} = (0,0,0,1)$ respectively. The Killing field 
$({{\partial}\over{\partial t}})^{a}$ is timelike for large values of $r$ but
becomes spacelike within the ergosphere whose position is given by: 
${r_+} \leq r \leq M + \sqrt{(M^2 - {a^2}~{{\cos^2}\theta})}$  
and is situated in the near-horizon region. 
This indicates that we can expand the solutions of the scalar field wave equation in terms of the eigenfunctions of the Killing vector fields. The eigenvalues of the Killing field 
$({{\partial}\over{\partial t}})^{a}$ can be interpreted as energy at a distance far away from the horizon.

The wave equation of a minimally coupled massless scalar field in a curved space is given by the following expression:

\be 
{(-g)^{-{1\over 2}}}{\partial_{\mu}}
[{(-g)^{{1\over 2}}}{g^{\mu \nu}}{\partial_{\nu}}
{\psi}] = 0 ~.
\ee

The wave equation is separable. As discussed above, we can take the basis function to be of the following form [15]:

\be 
f(l,m,p|x) = N(p){(r^2 + a^2)^{- 1/2}}{R_{lm}(p,a|r)}
{S_{lm}(aE|{\cos\theta})}{e^{im\phi}}{e^{-iEt}} ~. 
\ee

\noindent{Here $N(p)$ is a normalization constant, 
$p = E - m{\Omega_{H}}$ and 
${\Omega_{H}} = {a \over {{r_{+}}^{2} + a^2}}$ is the angular velocity of the event horizon.
$E$ is the energy measured by the asymptotic observer. $S_{lm}$ is a spheroidal harmonic satisfying the following equation:}

\be 
{\Big[}{d \over {d\xi}}(1 - \xi^2){d \over {d\xi}}
- {{m^2} \over {1 - \xi^2}} + 2maE - 
{(aE)^2}({1 - \xi^2}) + {\eta_{lm}(aE)}{\Big ]}
{S_{lm}(aE|{\xi})} = 0 ~.
\ee 

\noindent{The solutions can be expressed in terms of the oblate spheroidal functions [25] provided we set:}

\be 
{\eta_{lm}(aE)} = (aE)^2 - 2maE + \lambda_{lm}(aE) ~.
\ee

\noindent{Here $l = 0,1,2...$ and $m = -l, -l + 1, .., l - 1, l$.
The eigenvalues $\lambda_{lm}(aE)$ depend in a nontrivial way on $l, m, aE$ and is given by the following expression:}

\be 
\lambda_{lm}(aE) = {\sum_{k = 0}^{\infty}}{(-1)^k}{n_{2k}}~{(aE)^{2k}} ~.
\ee

\noindent{Here the sum is over $k = 0,1,2,...$ In the following we will only need the leading order terms in $l$. These are given by,}

\be 
{n_0} = l(l + 1); ~~~~~ 
{n_2} =  {1 \over 2} {\Big{[} 1 - 
{(2m - 1)(2m + 1) \over{(2l - 1)(2l + 3)}} \Big{]}} ~. 
\ee

The radial equation for ${R_{lm}(p,a|r)}$ is given by:

\be 
\Big{[}{{d^2} \over {d{r^*}^2}} - V_{lm}(E,a|r) \Big{]}
{R_{lm}(p,a|r)} = 0 ~.
\ee

\noindent{In the following we will be considering the near-horizon region of a macroscopic black hole and scalar waves with high values of $l$. In this limit, the function $V_{lm}$ is given by the following expression:}

\be 
V_{lm}(E,a|r) = - (E - m{a \over {r^2 + a^2}})^2 + 
{\eta_{lm}(aE)}{\Delta \over {(r^2 + a^2)^2}} ~.
\ee

\noindent{One can look at [12,15] for the details. Here
$r^*$ is the Regge-Wheeler tortoise coordinates in the 
equatorial plane and is given by the following expression:}

\be 
r^* = r + {2M \over {r_+ - r_-}} 
({r_+} ln{{r - r_+} \over {r_+}} - 
{r_-} ln{{r - r_-} \over {r_-}}) ~.
\ee

\noindent{The functions $V_{lm}$ take the values 
$-(E - m{\Omega_{H}})^2$ in the limit $r \rightarrow {r_+}$
and $-E^2$ in the limit $r \rightarrow \infty$. It acts as a potential barrier in the intermediate region. We find that we can construct solutions of the form 
$\sin[pr^* - \delta(p)]$ with:}

\be 
p = E - m{\Omega_{H}} ~.
\ee

\noindent{These solutions can be made to vanish at the brick wall in the near-horizon region with an appropriate choice of $\delta(p)$.}

Without any boundary conditions, $E$ has a continuous spectrum starting from zero and $l$ takes all positive integral values starting from zero. This leads to divergences when we try to evaluate the free energy and the entropy. As mentioned in Section:I, when the scalar field is in thermal equilibrium with the black hole we can use the brick wall model condition to evaluate the scalar field entropy. In the near-horizon region, the radial wave function takes the following form:

\be 
{R_{lm}(p,a|r)} \sim {\exp[{i\int}{k(r) dr}]} 
= {\exp[{i\int}{K(r^*) d{r^*}}]}
\ee

\noindent{Here the $r$-dependent radial wave vector is given by the following expression:}

\be 
k(r) = {\Big(}{{r^2 + a^2} \over \Delta}{\Big)}
\Big \{ (E - m{a \over {r^2 + a^2}})^2 - 
{\eta_{lm}(aE)}{\Delta \over {(r^2 + a^2)^2}} \Big \}^{1 \over 2} ~.
\ee 

\noindent{This indicates that we can use the WKB approximation to the wave equation. We found earlier [12,13] that in the spherical polar coordinates, the WKB approximation is suitable for a thin shell of scalar field. Since the near-horizon region gives us interesting physics, we proceed to evaluate the entropy of a thin shell of scalar field confined in the near-horizon region. We impose the brick wall boundary condition:}

\be 
\psi = 0, ~~~~ for  ~~~~r \leq {r_+} + h ~.
\ee 

\noindent{Thus, we can consider the scalar field to be confined in a half-infinite potential well in the near-horizon region. To implement the thin shell boundary condition, we consider solutions which are vanishing at the brick wall and are stationary up to a certain value 
$'d'$ of the radial coordinate. We then apply the WKB quantization rule. The brick wall boundary condition can be implemented by using linear combination of $R_{lm}$ and its complex conjugate. We have the following condition:}

\be 
h << d << {r_+} ~.
\ee

\noindent{We note that we can replace the brick wall boundary condition by a finite value. We can consider a 
steeply but smoothly rising potential in place of the brick wall. The WKB quantization rule obtained from the smooth matching condition will reproduce the same expressions for the thermal variables considered below. We keep the brick wall boundary condition because there exist solutions which are vanishing at the horizon [9] and due to reasons mentioned at the introduction. 
In the above, $h,d$ are coordinate variables and we will replace them by the corresponding proper distances. 
We have to be careful since the proper radial distance depends on the polar angle $\theta$.
The corresponding solutions are again stationary at a large value of the radial coordinate. The detail of the WKB approximation for the Schwarzschild black hole is discussed in [12].}

The radial quantum number is given by the WKB quantization rule:

\be
\pi n_d = {\int_{r_+ + h}^{r_+ + d}}{dr}~k(r,l,E) ~.
\ee

\noindent{To evaluate $n_d$, we have to assure that the wave number is real throughout the range of the radial integral
for each value of $E, l, m$ present in it's argument. This imposes constraints on the allowed values of $l$. In the Schwarzschild case $k(r)$ is independent of the azimuthal quantum number $m$ and we have the following expression for $k(r)$:}

\be
{k_{SCH}}(r) = {1\over{V(r)}}{\sqrt{{E^2} - {V(r)\over{r^2}}l(l + 1)}}~; ~~~~~
{K_{SCH}}(r^*) = {\sqrt{{E^2} - {V(r)\over{r^2}}l(l + 1)}}
~.
\ee
     
\noindent{Here, $V(r)$ is the metric function of the Schwarzschild black hole. 
The reality of $k(r)$ gives the following upper limit on the allowed values of $l$ [12,13]:}

\be
L(L + 1) = {{{E^2}{(2M + d + h)^3}} \over {d + h}} \approx
{{8{E^2}{M^3}} \over {d}} ~.
\ee

\noindent{In the present case, we have the following expression of $k(r)$ in the near-horizon region:}

\be 
k(r) = {\Big(}{{r^2 + a^2} \over \Delta}{\Big)}
\Big \{ p^2 - 
{\eta_{lm}(aE)}{\Delta \over {(r^2 + a^2)^2}} \Big \}^{1 \over 2} ~.
\ee 

\noindent{Here $p$ is given by Eqn.(13). The $r$ dependent term in the second parentheses is zero at the horizon and is monotonically increasing function of $r$ near the horizon.
Thus, we determine the maximum allowed value of $l$ by making $k(r) = 0$ at $r = d$ for a fixed $p$. This assures that $k(r)$ remains real in the intermediate region $(r_+ + h) \leq r \leq (r_+ + d)$ for all $l$ less than the maximum value of $l$ with a given $p$. The actual solutions, vanishing at the brick wall, are sinusoidal functions of $r^*$ obtained from the linear combination of $R_{lm}$ and its complex conjugate
with $E$ determined from Eqns.(15) and (18).}

In general it is difficult  to determine the allowed values of $l$ for a given $p$. We will use a few approximations which are appropriate for the WKB approximation in the near -horizon region. Since the WKB approximation holds good for high values of the quantum numbers, we can consider leading order terms in $l,m$. It is obvious that very near the horizon, $\Delta \approx 0$. In this case we can neglect the $r$-dependent term in Eqn.(21) and the allowed values of $l$ becomes free. It can be found easily that with a finite value of $E$, we can neglect terms with $k > 2$
in Eqn.(8) [25]. On the other hand, high values of $E$ are exponentially suppressed when we evaluate the partition function. We will consider the maximum possible value of ${\eta_{lm}(aE)}$ for a given large value of $l$ and this leads us to take $n_{2} = 0$ in Eqn.(8). We now have the following expression for $k(r)$:

\be 
{k}(r) = {\Big(}{{{r_+}^2 + a^2} \over \Delta(r)}{\Big)}
{\Big [}{p^2} - {{\Delta(r) {\Omega_H}^2} \over {a^2}}
\Big \{ {l(l + 1)} - 
2ma(p + m{\Omega_H}) +  {a^2}(p + m{\Omega_H})^2 \Big \} 
{\Big ]}^{1 \over 2} ~.
\ee

\noindent{Here we have taken $\Omega (r) = \Omega_H$.
We want to consider the maximum value of the second term
in the third parentheses with a given value of $p$ for 
$(r_+ + h) \leq r \leq (r_+ + d)$. As mentioned above, the $r$-dependent term is maximum at $(r_+ + d)$.
Since $a{\Omega_H} < {1 \over 2}$ for a non extreme black hole, it is appropriate to take $m = -l$ in the above expression and keep $p$ fixed by choosing appropriate $E$.
However, we will consider the leading order terms in $l$ and we have the following constraint on the allowed values of $l$:}

\be 
L^2 \leq {{a^2 p^2} \over {\alpha {\Delta(d)} {\Omega_H}^2}} ~.
\ee

\noindent{ Where,}

\be 
\alpha = {[1 - (a{\Omega_H})]^2} ~.
\ee

\noindent{For a Kerr black hole, $M \geq a$, 
$a{\Omega_H} \leq {1 \over 2}$ and the right hand side is never zero.   
Thus compared to the Schwarzschild black hole, the effect of rotation is to introduce the factor $\alpha$ in the leading order maximum allowed values of $l$. We will find the situation to be different when we calculate the scalar field entropy.}

We now turn to the entropy calculation. We will follow the following method developed in [7,12]. The stationarity spread of $\psi$ increases with the radial quantum number $n$. For the present purpose we consider the maximum value of $n$ given by Eqn.(18), multiply it by the corresponding angular degeneracy factor and integrate over $m,l,r,E$ respectively to obtain the free energy. This amounts to taking the same degeneracy factor for all the lower values of $n < n_d$. This method gives a good estimate of the entropy. This is evident from the nature of te near-horizon solutions and the final expression of the entropy.
The details of the method are given in [12]. 
Note that $n$ increases with $d$ while $L^2$ varies as
${{p^2} \over {\Delta(d)}}$. The radial parameter $d$ may be taken as the thickness of a thin shell in the near horizon region. We will fix the value of $d$ later.
The free energy is given by the following expression:

\ba
{\pi}{\beta}F & = & - {{{\int}_{0}^{\infty}}{{{\beta}~dE}\over{[{\exp}{ \{ {\beta (E - m{\Omega_H}) \}} - 1]}}}} \\ \nonumber
& ~ & {{\int_{r_+ + h}^{r_+ + d}}{dr}
~{\Big(}{{{r_+}^2 + a^2} \over \Delta(r)}{\Big)}} 
~{\int_{0}^{L} dl} 
~{\int_{-l}^{+l}dm}
~\Big \{ (E - m{\Omega_H})^2 - 
{\eta_{lm}(aE)}{\Delta(r) \over {({r_+}^2 + a^2)^2}} \Big \}^{1 \over 2} ~.
\ea

\noindent{As usual for the Kerr black hole, we have superradiant modes corresponding to $p = (E - m{\Omega_H}) < 0$. Thus in the above integral we replace $E$ by $p$ with the range of the $p$ integral from some large negative value to a high positive value of $p$. One should check that the constraint given by Eqn.(23) is still consistent. We have thus,}

\ba
{\pi}{\beta}F & = & - {{\int_{-P_0}^{\infty}}{{{\beta}~dp}\over{[{\exp}{ \{ {\beta p \}} -
1]}}}} \\ \nonumber
& ~ & {{\int_{r_+ + h}^{r_+ + d}}{dr}
~{\Big(}{{{r_+}^2 + a^2} \over \Delta(r)}{\Big)}} 
~{\int_{0}^{L} dl} 
~{\int_{-l}^{+l}dm}
~\Big \{ p^2 - 
{\eta_{lm}(p)}{\Delta(r) \over {({r_+}^2 + a^2)^2}} \Big \}^{1 \over 2} ~.
\ea

\noindent{Note that the Jacobian for the $(E,m) \rightarrow
(p,m)$ transformation is one.
We will discuss the superradiant modes and the lower limit later. We now concentrate on the $m$-integral. We consider only the quadratic terms in $m$ in the radical. This holds good even for $l \sim 10$. The $m$-integral then reduces to the following expression:

\ba  
{\int_{-l}^{+l}dm}~\sqrt{{A^2}{m^2} + B^2} & = & 
{l}{\sqrt{{A^2}{l^2} + B^2}} + {{B^2}\over{2A}}
~ \ln{\Big |}{{\sqrt{{A^2}{l^2} + {B}^2} + Al}
\over {\sqrt{{A^2}{l^2} + {B}^2} - Al}}{\Big |},
~ B^2 \geq 0 
\\ \nonumber
& = & {l}{\sqrt{{A^2}{l^2} - {B'}^2}} 
- {{{B'}^2}\over{2A}}
~ \ln{\Big |}{{\sqrt{{A^2}{l^2} - {B'}^2} + Al}
\over {\sqrt{{A^2}{l^2} - {B'}^2} - Al}}{\Big |}
,~ B^2 < 0 
\ea

\noindent{Here $A^2 = {a \Omega_H}(2 - {a \Omega_H})
{{\Delta{\Omega_H}^2}\over {a^2}}$ and 
$B^2 = -{B'}^2 = p^2 - {{\Delta{\Omega_H}^2}\over {a^2}}l^2$. Note that with $L$ given by Eqn.(23), $B^2$ can be negative.
We first consider the $a \rightarrow 0$ Schwarzschild limit. In this case $\alpha$ in Eqn.(24) can be taken to be one and $B^2$ can be taken to be positive. The right hand side then reduces to the Schwarzschild case expression:
${2l}{\sqrt{{E^2} - {V(r)\over{r^2}}{l^2}}}$ and the corresponding entropy will give the Schwarzschild case expression.}

We now consider the Kerr black hole with finite $a$.. It is evident from Eqn.(25) that the contribution of particles with negative values of the azimuthal quantum is exponentially small compared to that for the particles with positive values of the same. We should also consider the small value of the Hawking temperature for a macroscopic black hole. This makes the contribution of the negative values of $m$ negligible for a positive value of ${\Omega_H}$. 
This is consistent with the classical frame dragging in the near-horizon region. In the quantum case, the thin shell confined in the near-horizon region and in thermal equilibrium with the black hole preferentially contain particles whose azimuthal angular momentum is in the same direction as that of the horizon. 
This aspect is apparently lost in Eqn.(26) where everything is expressed in terms of the parameter $p$. However
it is evident from Eqn.(27) that 
the $m$ integral leads to the following expression in the leading order:

\be 
I(L) = {\int_{0}^{L} l~dl}
~\sqrt{{p^2} - {{{\alpha \Delta(r){{\Omega_H}^2}} \over {a^2}}l^2}}  ~. 
\ee

\noindent{It is useful to note that the radicals in Eqn.(27) is small compared to $l$ for high values of $l$.
When we compare this expression with the Schwarzschild case, we find that the factor $(2l + 1)$ is replaced by $l$.
Thus, the final expression of the entropy will be half of the corresponding expression in the Schwarzschild black hole. This is consistent with the preferential emission of the scalar particles with positive azimuthal quantum number.
Note that the lower limit of the $l$ integral is taken to be zero. Lower order terms neglected in Eqn.(11) become important in this limit [15]. This will give a negligible contribution since the upper limit is too high.}

\noindent{We now break the energy integral into two parts. The first part corresponds to the ordinary modes and have $p \geq 0$.
The second part contains the superradiant modes with 
$p < 0$. We first evaluate the free energy for the ordinary modes. The calculation remains similar to the Schwarzschild case with an additional half factor.  We have the following expression for the free energy:}

\ba
{\pi}{\beta}F & = & - {{\int_{0}^{\infty}}{{{\beta}~dp}\over{[{\exp}{ \{ {\beta p \}} -
1]}}}} \\ \nonumber
& ~ & {{\int_{r_+ + h}^{r_+ + d}}{dr}
~{\Big(}{{{r_+}^2 + a^2} \over \Delta(r)}{\Big)}}
~{1 \over 3} {\Big [} p^3 - 
\{p^2 - 
{{{s \alpha \Delta(r){{\Omega_H}^2}} \over {a^2}}L^2}\}^{3/2}
{\Big ]} ~.
\ea

\noindent{Here $s$ is a parameter $\leq 1$. 
We can now do a binomial expansion in the radial integrand in terms of $s$. We will consider only the linear order term in $s$ since $\Delta(r)$ is small in the near-horizon region. The higher order terms for the Schwarzschild black hole is given by Eqn.(35) of [12].  
The trial solution given by Eqn.(14) will satisfy Eqn.(10) provided we can neglect the ${{dK(r^*)}\over {dr^*}}$ term. Note that the actual solutions are sinusoidal functions discussed below Eqn.(21).  
In the near-horizon region ${dK(r^*)}\over {dr^*}$ is proportional to $(1 - {{2M \over r}})$ for the Schwarzschild black hole and we can neglect this term compared to $V_{lm}$ given by Eqn.(11). Similar aspect will remain valid for the Kerr black hole with $(1 - {{2M \over r}})$ replaced by ${{\Delta(r)} \over {({r_+}^2 + a^2)}}$. Thus, there is no additional constraint on the allowed values of $l$ apart from Eqn.(20) and we can put $s= 1$ in the final expressions for both the Schwarzschild and Kerr black hole. This will change the numerical estimate for the brick wall cut-off given in [12] by insignificant amount. We had taken $s = {10^{-2}}$ in [12]. We will mention the corresponding values in Section:III. However, it is helpful to keep $s$ as an expansion parameter in the intermediate steps. This helps us to identify terms of different orders in $\Delta(r)$ easily.
We previously used $s$ to implement the WKB condition:
${|{\nabla[S(r)]}|^2} >> |{\nabla}^2 [S(r)]|$ on the solutions of the form $\rho(r){\exp[{iS(r)}}]$
which are stationary throughout the manifold. This gives
the constraint: $s << 1$ for the globally stationary solutions in the Schwarzschild black hole [12]. 
This can be found from Eqns.(10,14,19) with $r = 3M$. The significance of the point $r = 3M$ can be found from [8,12].} 

The leading order term of free energy is given by the following expression:

\be 
F = -{1 \over \beta^4}~{{\pi^3} \over {30}}
~{{({r_+}^2 + a^2)^3} \over {(r_+ - r_-)^2}}
~{1 \over d}~{ln({d \over h})} ~.
\ee

As expected, the free energy is proportional to the fourth power of the temperature. We now introduce the covariant cut-off parameter defined by,

\be 
{\epsilon_b} = {\int_{r_+}^{r_+ + b}}
{\rho \over {\sqrt{\Delta(r)}}}~dr ~.
\ee

\noindent{It is obvious that the proper radial distance depends on $\theta$. We have the following relation between the proper radial distance at the point where the ergosphere vanishes (here $\theta = 0$ or $\pi$) and at an arbitrary value $\theta$ of the polar angle:}

\be 
{{\epsilon_\theta}^2} = 
(1 - {a \Omega_H}~{{\sin^2}\theta}){\epsilon^2} ~.
\ee  

\noindent{Here $\epsilon$ is the proper radial distance along $\theta = 0$ or $\pi$.  With ${a{\Omega_H}} < {1 \over 2}$, different choices of $\theta$ lead to a minor numerical factor. However, the expression of the free energy given by Eqn.(30) is independent of $\theta$.
Note that the ergosphere is covariantly defined in terms of the invariant quantities: the norm of the Killing field $({{\partial}\over{\partial t}})^{a}
+ {\Omega_H}({{\partial}\over{\partial \phi}})^{a}$
and the norm of the Killing field $({{\partial}\over{\partial t}})^{a}$.
We choose the radial brick wall cut-off and the radial thickness in such a way that they yield definite values for the corresponding proper variables along the directions where the ergosphere vanish. Thus we may consider that we have a thinshell in the near-horizon region whose boundaries have ${\theta}$-dependent proper radii as measured from the horizon. The free energy and the entropy are expressed in terms of the proper radial variables evaluated along the directions along which the ergosphere vanish. In the familiar Kerr metric given by Eqn.(1), the polar axis represent the points where the ergosphere vanish. 
\textit {In all the subsequent discussions, $\epsilon$ will refer to the proper radial distance determined in the this way.} We will discuss the numerical values of the proper variables in Section:III.}

The entropy can be obtained from the relation:
$S = {\beta^2}{{dF \over {d\beta}}}$. In terms of the covariant parameters the leading order expression is given by:

\be 
S = {{16 \pi^3} \over {15}}~{1 \over \beta^3}
~{{({r_+}^2 + a^2)^4} \over {(r_+ - r_-)^3}}
~{1 \over {\epsilon_d}^2}~{ln({{\epsilon_d} \over{\epsilon_h}})} ~.
\ee

\noindent{For the temperature, we take the value measured
by the asymptotic static observer at infinity:}

\be 
\beta = {{4\pi({r_+}^2 + a^2)} \over(r_+ - r_-)} ~.
\ee

\noindent{The entropy of a scalar field confined within the near horizon region and predicted by an observer at infinity is then given by,}

\be
S = {{1} \over 60}~{ln({{{\epsilon}_d} \over {{\epsilon}_h}})}~{A \over {4 \pi {{\epsilon_d}^2}}} ~.
\ee   
    
\noindent{The corresponding expression in the Schwarzschild black hole is given by the following expression:}

\be
S' = {{1} \over 30}~{ln({{{\epsilon}_d} \over {{\epsilon}_h}})}~{A \over {4 \pi {{\epsilon_d}^2}}} ~.
\ee 

\noindent{As mentioned earlier, the leading order entropy of the scalar field in the Kerr black hole is half of the corresponding value in the Schwarzschild black hole.} 

We now turn to the superradiant modes. The corresponding free energy is given by the following expression:

\ba
{\pi}{\beta} F' & = & - {{\int_{-P_0}^{0}}{{{\beta}~dp}\over{[{\exp}{ \{ {\beta p \}} -
1]}}}} \\ \nonumber
& ~ & {{\int_{r_+ + h}^{r_+ + d}}{dr}
~{\Big(}{{{r_+}^2 + a^2} \over \Delta(r)}{\Big)}} 
~{\int_{0}^{L} dl} 
~{\int_{-l}^{+l}dm}
~\Big \{ p^2 - 
{\eta_{lm}(p)}{\Delta(r) \over {({r_+}^2 + a^2)^2}} \Big \}^{1 \over 2} ~.
\ea

\noindent{The constraint given by Eqn.(23) and $E \rightarrow \infty$ indicate that ${P_0} \rightarrow \infty$. Note that we always have $E \geq 0$.  
However, with $p$ being negative, the thermal factor becomes approximately $-1$ for all $p$ and the integral diverge badly. We can keep $P_0$ finite and take the limit ${P_0} \rightarrow \infty$ at the end. In this case the free energy becomes independent of the temperature and the entropy vanishes. This is not unexpected since for the superradiant modes, the stimulated emission dominates and the thermal radiation is not much significant. Thus we exclude the superradiant modes from our consideration and regard Eqn.(35) to be the leading order entropy of a minimally coupled scalar field confined in the near horizon region and in thermal equilibrium with the Kerr black hole.}

\section{Discussion}

Equations (35) and (36) give the leading order expressions for the entropy of the scalar field in the Kerr and Schwarzschild black hole respectively. It is easy to find higher order terms. In the Schwarzschild black hole the higher order terms in $\frac{V(r)}{r^2}$ are again proportional to the area but are negligible compared to the expression given by Eqn.(36). This is given by Eqn.(35) of [12]. The degrees of freedom of a three dimensional thin shell in the near-horizon region is mostly related to the inner boundary which is almost coincident with the horizon. Similar result will also remain valid for the Kerr black hole. This is similar to the black hole entropy itself. Here we consider another sub-leading term in the Schwarzschild black hole associated with linear order in $\frac{V(r)}{r^2}$, [12]. Corresponding entropy for a massless scalar field is given by:

\be
S = \frac{4 \pi^3}{15d} (2M) {(\frac{2M}{\beta})^3} ln(\frac{d}{h})  + \frac{4 \pi^3}{15 \beta^3} (\frac{d}{d + h}) (2M + d + h)^3
\ee

\noindent{For finite $M$ and $\beta = 8 \pi M$, the second term is small compared to the first term. However, in the flat spacetime limit: $M \rightarrow 0$, and $h \rightarrow 0$ the second term gives the most significant contribution and is proportional to $(\frac{d}{\beta})^3$, where $d$ is now the radius of a small sphere. This is similar to the corresponding flat spacetime expression $\sim \frac{volume}{\beta^3}$. In Kerr black hole, the flat spacetime limit is obtained by considering $M \rightarrow 0, a \rightarrow 0$ and $h \rightarrow 0$.}

A related field is to determine the expressions for the entanglement entropy
of a scalar field in a given spacetime. The 
entanglement entropy of the matter fields are found from the reduced density
matrices obtained by summing over the degrees of freedom confined within 
a given spatial region of the flat space time
[26,27,28]. 
This is significant for a scalar field in the near horizon region of both the Schwarzschild and the Kerr black hole. In terms of the tortoise coordinates,
the wave equation is similar to that of a one-dimensional free particle
in the flat space when we consider the near-horizon region of both the black holes. In the Kerr black hole we have to consider the classical frame dragging and it's consequences for the quantum case. This is obvious from Section:II. 
We will later discuss this issue. In this context, it will be interesting to consider the entanglement entropy for a thin shell of scalar field surrounding the horizon and compare the result with the expression obtained in this article. This may be important to explain the black hole entropy in terms of the near horizon states.

The expression of the scalar field entropy given by Eqn.(35)
contains two free parameters, the proper thickness of the thin shell and the proper brick wall cut-off parameter
as defined in Section.II. We first fix the proper radial thickness of the thin shell. In the Schwarzschild black hole, we took the thickness
of the thin shell to be of the order of atomic lengths [12]. Here also we choose the same. The other undetermined parameter is the proper brick wall cut-off. It had been shown by Candelas
that the stress tensor of the scalar field in the Hartle-Hawking vacuum is finite [16]. 't Hooft wanted to explain the black hole entropy in terms of the near horizon part of the matter field entropy and equated the two [7]. If we do so, the internal energy turns out to be finite: $U = {3 \over 8}M$.
This also determines the proper brick wall cut-off parameter. In the present case we have:

\be
{\epsilon_h} = {{\epsilon}_d}
{\exp[-{60 \pi}~{{({{\epsilon_d} \over {l_p}})}^2}]} ~.
\ee 

\noindent{Where $l_p$ is the Plack length. Note that this is different from the Schwarzschild value [12]:}

\be
{\epsilon_h} = {{\epsilon}_d}
{\exp[-{30 \pi}~{{({{\epsilon_d} \over {l_p}})}^2}]} ~.
\ee

\noindent{If we take
${\epsilon_d} = {10^{-10}}~cm$, $\epsilon_h$ is given by the following expression in terms of the Plack length:}

\be
{\epsilon_h} \sim [{{10}^{24}}]{\exp({-{{10}^{50}}})} ~.
\ee

\noindent{Thus, the brick wall is almost coincident with the horizon although the free energy is finite. This is expected since there exists solutions vanishing on the horizon. Note that the power $50$ in the above equation is replaced by ${{50} \over s}$ for $s < 1$. For any $s < 1$, this lowers the value of ${\epsilon_h}$ further. 
We now consider the internal energy of the thin shell. When $\epsilon_h$ is given by Eqn.(38), the internal energy is given by: $U = {3 \over 8}~{\sqrt{M^2 - a^2}}$.
Alternatively, we can keep the value of the proper brick wall cut-off same as that for the Schwarzschild black hole.
In this case the entropy of the thin shell will be half of that of the black hole and the internal energy is given by:
$U = {3 \over 16}~{\sqrt{M^2 - a^2}}$. In both the cases
the internal energy is finite and we may disregard the back reaction problem upto a first approximation.}

In passing, we note that the WKB approximation is a first approach. It gives expected result for a thin shell of matter field. We considered the almost vanishing transmission coefficient for the modes which are only stationary in the near-horizon region. They are again stationary at a large distance away from the horizon with a vanishingly small amplitude. These aspects lead us to consider the entropy of a thin shell of matter field confined in the near-horizon region. We obtain expected results for the Kerr-Newman family of black hole. This is consistent with the nature of the scalar wave equation in terms of the tortoise coordinates. In the near-horizon region, this is similar a one-dimensional free wave equation in the flat space to a first approximation.}

The brick wall model approach is based on the Lorentzian sector of the black hole
space time and is expected to give the most robust expression. 
It is convenient to use the Zeta function regularization
scheme in the Euclidean sector calculation for the partition function
of a scalar field in a black hole background. This is considered in [18,29,30] for the Schwarzschild black hole with the
boundary conditions similar to those used in this article.
In the Zeta function regularization scheme to evaluate the partition function, 
the partition function is given by the Eqn(3.2) of [16]:

\be
\ln[Z] = {1 \over 2}[{\zeta}'(0) + \ln({1 \over 4}{\pi}{{\mu}^2}){\zeta}(0)] ~.
\ee

\noindent{Here $\mu$ is a normalization factor and $\zeta(s)$ is the Zeta function. The evaluation of the scalar field entropy is considered by many authors. One can look at [17] for a review. The WKB approximation is also used here [15]. 
The expression of entropy contains an undetermined parameter. The expression of the entropy obtained by us can be used to determine the undetermined parameter in terms of the physical brick wall parameter. The logarithmic divergence obtained by us is significant in this respect. It renders quadratically and quartically divergent terms in the Euclidean sector expression to logarithmically divergent. 
The Euclidean sector calculation of the scalar field entropy in the Kerr black hole is not well formulated [20,21]. The expression obtained by us in this article is thus illuminating in this context.}

The scalar field entropy have been interpreted by some authors as a quantum correction
to the black hole entropy and thus giving infinite renormalization
to the bare gravitational constant $G_{B}$. A few authors used the brick wall model for this purpose [31,32,33]. 
The one-loop effective action of a
scalar field in a curved space-time can be found by using the DeWitt\textendash Schwinger proper time representation.
The divergent parts are given by equation (6.44) in [34]. These terms are divergent only when
the space-time dimension is four and may be interpreted as giving infinite renormalizations   
to the different coupling constants present in the Einstein-Hilbert action for the gravitational
field itself. The divergent term is of the form $1 \over {(n - 4)}$, where $n$ is the space time
dimension. From Eqns. (35) and (36) it is obvious that the scalar field entropy in the Schwarzschild and Kerr black hole are logarithmically divergent. Similar aspect also remains valid in Reissner-Nordstrom black hole. Thus, we may not fix the brick wall cut-off parameter as we have done before. We add the divergent part of scalar field entropy to the black hole entropy with $G$ replaced by $G_B$ and absorb the divergence to give a 
renormalized value. The structure of this quantum correction is different from a quadratically divergent expression obtained before. Moreover, this scheme can work if we find that the matter field entropies in different black holes are of similar form as obtained for the asymptotically flat black holes considered here. Thus, they should contain a term proportional to the gravitational entropy. The exception can be the extremal black holes for which the black hole entropy is vanishing. In this context it will be interesting to use the methods developed by us to calculate the entropy of a minimally coupled scalar field in non-trivial black holes like the Taub-NUT space time and the constant curvature black holes.

\section{Conclusion}

To conclude, in this paper we have found out the entropy of a minimally coupled scalar field in the Kerr black hole background. We have used the Lorentzian sector brick wall  model of 't Hooft. We have used an improved method to calculate the entropy previously developed by us. We have considered a thin shell of scalar field confined in the near-horizon region and in thermal equilibrium with the black hole. The leading order term of corresponding entropy is proportional to the area of the event horizon and is logarithmically divergent in a covariant radial parameter known as the proper radial brick wall cut-off parameter. This is an improvement compared to the earlier results which are quadratically divergent in the  
proper radial brick wall cut-off parameter. The logarithmic divergence is expected due to gravitational redshift.  
The method used by us is significantly different from that used by 't Hooft. 
For a fixed value of the brick wall cut-off, the scalar field entropy in the Kerr black hole is found to be half of the corresponding expression in the Schwarzschild black hole background. 
This is expected for the thermal radiation in the Kerr black hole due to preferential emission of particles with
azimuthal angular momentum in the same direction as that of the black hole. The contribution of the
particles in the thin shell with opposite values of the azimuthal angular momentum is exponentially
suppressed. This is due to rotation of the black hole. However, we can obtain the Schwarzschild case expression
by including a subleading term and taking the appropriate limit. We have briefly discussed the
significances of the results in Sect.3. We have also discussed a subleading term for the Schwarzschild black hole that gives an expression similar to that in flat spacetime in the limit $M \rightarrow 0$. Thus, the method used by us give expressions that are consistent with both gravitational blueshift and flat spacetime limit. Further significance with regard to the entanglement approach to explain the black hole entropy and the holographic principles will be discussed later.

\section{References}

[1] J. M. Bardeen, B. Carter and S. W. Hawking, Commun. Math. Phys. {\bf 31}, 161 (1973).

[2] J. Bekenstein, Phys. Rev. D {\bf 7}, 2333 (1973). 

[3] S. W. Hawking, Commun. Math. Phys. {\bf 43}, 199 (1975).

[4] S. A. Fulling, Phys. Rev. D {\bf 7}, 2850 (1973). 

[5] W. G. Unruh, Phys. Rev. D {\bf 14}, 870 (1976).

[6] J. B. Hartle and S. W. Hawking, Phys. Rev. D {\bf 13},      

    2188 (1976).
    
[7] G. 't Hooft, Nucl. Phys. B {\bf 256}, 727 (1985).

[8] R. M. Wald, General Relativity (The University of Chicago Press, Chicago and London, 1984).

[9] D. G. Boulware, Phys. Rev. D {\bf 11}, 1404 (1975).

[10] I. Ichinose and Y. Satoh, Nucl. Phys. B {\bf 447}, 34 (1995).

[11] S. P. de Alwis and N. Ohta, Phys. Rev. D {\bf 52}, 3529 (1995).

[12] K. Ghosh, J. Phys. Soc. Japan {\bf 85}, 014101, (2016);  K. Ghosh, arXiv:0902.1601, v9 (2016).

[13] K. Ghosh, J. Phys. Conf. Ser. {\bf 410}, 012137 (2013).

[14] K. Ghosh, Nucl. Phys. B {\bf 814}, 212 (2009). 

[15] B. S. DeWitt, Phys. Reports {\bf 19}, No.6 (1975).

[16] P. Candelas, Phys. Rev. D {\bf 21}, 2185 (1980).

[17] S. N. Solodukhin, Living Rev. Relativity {\bf 14}, 8 (2011).

[18] S. W. Hawking, Commun. Math. Phys. {\bf 55}, 133 (1977).

[19] G. W. Gibbons and S. W. Hawking, Commun. Math. Phys. {\bf 66}, 291 (1979).

[20] V. Frolov and D. Furasarv, Class. Quant. Grav. {\bf 15}, 2041 (1998) 

[21] D. Fursaev and S. Solodukhin, 
     Phys. Rev. D {\bf 52}, 2133 (1995).

[22] G. Cognola, L. Vanzo and S. Zerbini, Class.Quant.Grav. {\bf 12}, 1927 (1995).

[23] A. A. Starobinsky, Sov. Phys.-J.E.T.P {\bf 37}, 28 (1973); R. Penrose, Rev. del Nuovo Cimento {\bf 1}, 
252, (1969).

[24] W. G. Unruh, Phys. Rev. D {\bf 10}, 3194 (1974). 

[25] M. Abramowitz and I. A. Stegun, Handbook of Mathematical Functions (Dover Publications, Inc., New York, 1964)

[26] L. Bombelli, R. K. Kaul, J. Lee and R. Sorkin, Phys. Rev. D {\bf 34}, 373 (1986).

[27] C. Callan and F. Wilczek, Phys. Lett. B {\bf 333}, 55 (1994).

[28] M. Saravani, R. D. Sorkin and Y. K. Yazdi, Class. Quant. Grav. {\bf 31}, 214006 (2014).

[29] D. B. Ray and I. M. Singer, Adv. in Math. {\bf 7}, 145 (1971).

[30] P. B. Gilkey, Adv. in Math. {\bf 15}, 334 (1975).

[31] J. G. Demers, R. Lafrance and R. C. Myers, Phys. Rev. D {\bf 52}, 2245 (1995).

[32] L. Susskind and J. Uglum, Phys. Rev. D {\bf 50}, 2700 (1994).

[33] A. Ghosh and P. Mitra, Phys. Lett. B {\bf 357}, 295 (1995).

[34] N. D. Birrel and P. C. W. Davies, Quantum Fields in Curved Space (Cambridge University Press, Cambridge, 1982).

\end{document}